\documentclass[conference]{IEEEtran}
\usepackage{cite}
\IEEEoverridecommandlockouts
\usepackage{cite}
\usepackage{amsmath,amssymb,amsfonts}
\usepackage{graphicx}
\usepackage{textcomp}
\usepackage{xcolor}
\usepackage{soul}
\usepackage{tabularray}
\usepackage{booktabs}
\usepackage{fancyhdr}
\usepackage{hyperref}
\usepackage{tabularray}
\usepackage{float}
\usepackage{booktabs}
\usepackage{xfrac}
\usepackage{algorithm,algpseudocode}
\usepackage{multirow}
\usepackage{amsmath, bm}
\usepackage{subcaption}

\renewcommand{\figurename}{Figure.}
\renewcommand{\tablename}{Table.}

\def\BibTeX{{\rm B\kern-.05em{\sc i\kern-.025em b}\kern-.08em
    T\kern-.1667em\lower.7ex\hbox{E}\kern-.125emX}}
\begin{document}

\title{CiFlow: Dataflow Analysis and Optimization of Key Switching for Homomorphic Encryption
}
\author{\IEEEauthorblockN{Negar Neda\IEEEauthorrefmark{1},
Austin Ebel\IEEEauthorrefmark{1},
Benedict Reynwar\IEEEauthorrefmark{2}, and 
Brandon Reagen\IEEEauthorrefmark{1}}
\IEEEauthorblockA{\IEEEauthorrefmark{1}New York University, New York, New York, USA, {\{negar, 
abe5240, bjr5\}@nyu.edu}}
\IEEEauthorblockA{\IEEEauthorrefmark{2}Information Sciences Institute, University of Southern California, Arlington, VA, USA, 
breynwar@isi.edu}}

\thispagestyle{plain}
\pagestyle{plain}
\maketitle

 \begin{abstract}
Homomorphic encryption (HE) is a privacy-preserving computation technique that enables computation on encrypted data. 
Today, the potential of HE remains largely unrealized as it is impractically slow, preventing it from being used in real applications.
A major computational bottleneck in HE
is the key-switching operation, accounting for approximately 70\% of the overall HE execution time and involving a large amount of data for inputs, intermediates, and keys.
Prior research has focused on hardware accelerators to improve HE performance, typically featuring large on-chip SRAMs and high off-chip bandwidth to deal with large scale data.



In this paper, we present a novel approach to improve key-switching performance by rigorously analyzing its dataflow.
Our primary goal is to optimize data reuse with limited on-chip memory to minimize off-chip data movement.
We introduce three distinct dataflows: Max-Parallel (MP), Digit-Centric (DC), and Output-Centric (OC), each with unique scheduling approaches for key-switching computations. 
Through our analysis, we show how our proposed Output-Centric technique can effectively reuse data by significantly lowering the intermediate key-switching working set and alleviating the need for massive off-chip bandwidth.
We thoroughly evaluate the three dataflows using the RPU, a recently published vector processor tailored for ring processing algorithms, which includes HE. This evaluation considers sweeps of bandwidth and computational throughput, and whether keys are buffered on-chip or streamed.
With OC, we demonstrate up to 4.16$\times$ speedup over the MP dataflow and show how OC can save 12.25$\times$ on-chip SRAM by streaming keys for minimal performance penalty.
\end{abstract}
\section{Introduction}
Today, many computations are outsourced to the cloud, whether to leverage its scale or overcome performance constraints of locally available devices (e.g., smartphones).
Despite the success of this computational model, it remains vulnerable to attack and does not provide users with strong privacy and security guarantees over who can view and use their data. 
Homomorphic encryption (HE)~\cite{gentry2009fully} offers a solution.
With HE, functions computed on ciphertext are also directly applied to the underlying plaintext.
This facilitates direct computation on encrypted data, extending cryptographic security from communication and storage to include computation for complete end-to-end secure outsourced computation.


Many HE schemes (e.g., BGV~\cite{brakerski2014leveled}, BFV~\cite{brakerski2012fully, fan2012somewhat}, and CKKS~\cite{cheon2017homomorphic}) and software implementations (e.g., OpenFHE~\cite{OpenFHE}, Lattigo~\cite{lattigo}, SEAL~\cite{sealcrypto}) now exist and have made steady performance improvements over the years. However, practical deployment of HE is still limited by significant performance overheads, attributed to large data sizes, complex operations, and additional functions to process.
When encrypting and processing HE ciphertexts, the data becomes significantly larger than plaintext, forming large ciphertext vectors of up to 2$^{17}$ elements.
Here, each plaintext element is represented by hundreds to thousands of bits, and special precomputed keys ($\mathbf{evks}$) are needed for certain operations (i.e., multiplication and rotation) that can be upwards of hundreds of MBs.
This puts substantial pressure on the memory system.
Next, HE requires complex modular arithmetic for processing each element, which is not natively supported by mainstream commercial processors.
Finally, HE operations are not one-to-one conversions of plaintext operators;
they involve multiple additional functions for performance (e.g., Number Theoretic Transforms (NTT)) and correctness before and after the actual operation (e.g., multiplication and rotation).
These factors collectively
overwhelm commodity hardware, and prior work has consistently reported 4-6 orders of magnitude slowdown compared to plaintext~\cite{reagen2021cheetah, samardzic2021f1, aikata2023reed, fan2022tensorfhe }.


The slowdown has been addressed by a growing body of work on hardware acceleration for HE.
Many solutions now exist that consider fixed-function pipelines~\cite{reagen2021cheetah, riazi2020heax}, vector architectures~\cite{samardzic2021f1, samardzic2022craterlake, soni2023rpu}, and tiled architectures~\cite{kim2022bts, kim2022ark}.
The architectural approaches taken differ, but the building blocks are the same: very large on-chip memories and high off-chip bandwidth. For example, CraterLake~\cite{samardzic2022craterlake}) uses 256 MB of on-chip SRAM and assumes two HBM2e PHYs for a total bandwidth of 1TB/s.
BTS~\cite{kim2022bts} and ARK~\cite{kim2022ark} both have 512 MB of on-chip SRAM and assume 1TB/s of memory bandwidth.
The large on-chip SRAMs and multiple PHYs result in large chips, adding the cost of the advanced memory technology results in expensive solutions.


Our hypothesis is that by optimizing HE dataflow, we can capture on-chip reuse with far less SRAM and simultaneously reduce the off-chip bandwidth requirements.
To understand the potential of optimizing dataflow, we deeply analyze the \emph{hybrid key-switching algorithm (HKS)}~\cite{han2020better}.
HKS is the core computation of HE and highly complex.
The use of HKS in HE is widespread; it is called after each rotation and homomorphic multiplication and is heavily used in bootstrapping.
For example, recent work has reported that a single HE ResNet-20 inference takes 3,306 rotations \cite{pmlr-v162-lee22e}, and prior research has further shown that HKS dominates HE's runtime~\cite{decastro2021does,kim2022ark, 10070953}, and can be up to 70\% for private neural inference~\cite{pmlr-v162-lee22e}.


A single HKS execution can involve hundreds of NTTs, hundreds of MBs of input and output data, nearly 500MB of constant $\mathbf{evks}$, and up to 1GB of intermediate data. Thus, existing solutions to processing the workload effectively rely on large on-chip SRAMs and high off-chip bandwidth.

To understand and optimize HKS, we propose three distinct dataflows: Max-Parallel (MP), Digit-Centric (DC), and Output-Centric (OC).
Our key insight is that with the OC dataflow, the intermediate state of HKS can be significantly compressed while maintaining high parallelism to utilize computational units.
To demonstrate OC's potential, we implement five parameterizations of HKS taken from recent work following each of the three presented dataflows on a recent vector HE accelerator~\cite{soni2023rpu}.
Using previously validated simulation infrastructure, we rigorously evaluate the dataflows.
First, we assume a large on-chip SRAM (i.e., 392 MB), which is sufficient to buffer all $\mathbf{evks}$ on-chip while reserving 32MB for data (i.e., inputs and intermediates).
We find that OC consistently outperforms other dataflows and can deliver up to 4.16$\times$ speedup over MP using the same bandwidth.
Next, given that $\mathbf{evks}$ have no reuse within HKS, we elect to stream them on-chip and reserve a fraction of off-chip bandwidth for them.
Here again, OC performs well.
With our OC dataflow, we can stream $\mathbf{evks}$ to reduce on-chip SRAM by 12.25$\times$ and bandwidth by 3.3$\times$, while achieving the same performance as an MP on-chip implementation.


In summary, our contributions are:
\begin{enumerate}
    \item We propose three different dataflows for HKS algorithm, named Max-Parallel, Digit-Centric, and Output-Centric. Through our analysis we demonstrate how strategically scheduling instructions and efficiently reusing loaded and generated data by the OC dataflow can lead to substantial off-chip bandwidth saving. 
    \item We evaluate two scenarios for handling $\mathbf{evks}$: streaming from off-chip and on-chip storage with a larger memory. Streaming $\mathbf{evks}$ reduces SRAM area by 12.25$\times$ and OC saves 3.3$\times$ bandwidth over the MP baseline.
    \item We evaluate the performance of dataflows across different off-chip bandwidths and computational throughput to understand the bandwidth-compute trade-off. We find OC to be highly effective, matching naive MP HKS implementation with significant bandwidth saving. Furthermore, increasing accelerator throughput enhances performance even further.
\end{enumerate}

\section{Background}
\label{sect:background}

In this section we briefly introduce the CKKS HE scheme \cite{cheon2017homomorphic}, including its key parameters and operations, using terminology from ARK\cite{kim2022ark} and Castro~\cite{decastro2021does} when applicable. Unlike prior HE schemes, BGV~\cite{brakerski2014leveled} and BFV\cite{brakerski2012fully, fan2012somewhat}, CKKS operates on vectors of real or complex numbers, rather than integers. To perform these operations, a vector message, $\mathbf{m}$, is first encoded into a plaintext $\textit{polynomial}$ $R_Q = \mathbb{Z}_Q[X] / (X^N + 1)$, represented as $[\mathtt{P}]$. In essence, this is another vector of length $N$, with each element being an integer no larger than Q. Depending on a problem's complexity, the value of $N$ typically ranges from $2^{11}$ to $2^{17}$, while $Q$ can range from hundreds to thousands of bits. For a polynomial with $N$ coefficients, a vector message must be of length $n \leq \sfrac{N}{2}$.

\setlength{\tabcolsep}{4pt}
\renewcommand{\arraystretch}{1.2}
\begin{table}
   \caption{Relevant CKKS Parameters.}
   \label{tab:params}
   \centering
   \begin{tabular}{cl}
   \toprule\toprule
   \textbf{Param.}                            & \textbf{Description} \\
   \midrule
   $\mathbf{m}$                               & Message, a vector of real or complex numbers. \\
   $[\![ \mathbf{m} ]\!]$, $[\![ \mathbf{m} ]\!]_s$ & Ciphertext encrypting the message, $\mathbf{m}$, under secret key $s$. \\
   $N$                                        & Power-of-two polynomial ring degree. \\ 
   $n$                                        & Length of the vector message, $n \leq \sfrac{N}{2}$. \\
   $Q$                                        & Initial polynomial modulus.\\
   $P$                                        & Auxiliary modulus used in key-switching. \\
   $L$                                        & Maximum multiplicative level of $[\![ \mathbf{m} ]\!]$. \\
   $\ell$                                     & Current multiplicative level and remaining \textit{towers}. \\
   $K$                                        & Number of moduli/towers in $P$. \\
   $q_i$                                      & Small moduli in RNS decomposition of $Q = \prod_{i=0}^{L} q_i$. \\
   $p_i$                                      & Small moduli in RNS decomposition of $P = \prod_{i=0}^{K-1} p_i$. \\
   $\mathtt{P}$                               & Polynomial in $\mathbb{Z}[X]/(X^N+1)$. \\
   $\mathcal{B_\ell}$, $\mathtt{[P]}_\mathcal{B_\ell}$    & The set of primes $\{q_0, q_1, \ldots, q_\ell\}$, $\mathtt{[P]}_{i \in \mathcal{B_\ell}}$.\\
   $\mathcal{C}$, $\mathtt{[P]}_\mathcal{C}$    & The set of primes $\{p_0, p_1, \ldots, p_{K-1}\}$, $\mathtt{[P]}_{i \in \mathcal{C}}$. \\
   $\mathcal{D_\ell}$, $\mathtt{[P]}_\mathcal{D_\ell}$    & The union of $\mathcal{B_\ell} \cup \mathcal{C}$, $\mathtt{[P]}_{i \in \mathcal{D_\ell}}$. \\
   $dnum$                                     & Number of \textit{digits} that $\mathtt{P}$ is decomposed into. \\
   $\alpha$                                   & The number of towers in each digit, $\lceil \sfrac{(L+1)}{dnum} \rceil$. \\
    $\mathbf{evk}$                             & Evaluation key used to convert $[\![ \mathbf{m} ]\!]_{s'}$ to $[\![ \mathbf{m} ]\!]_{s}$. \\
   \bottomrule
   \end{tabular}
   \vspace{-2em}
\end{table}
A ciphertext, $[\![ \mathbf{m} ]\!]$, consists of a pair of polynomials, $(C_0, C_1)$, where one polynomial contains the message with a small amount of random noise added to it to ensure the security of the RWLE scheme. For efficiency, ciphertexts are decomposed into an equivalent RNS representation \cite{cheon2019full} consisting of many smaller, machine-word size moduli, $q_i$, such that $R_{q_0} \times R_{q_1} \times \ldots \times R_{q_\ell} = R_Q$. Each small moduli typically ranges from $36$ to $64$ bits \cite{kimsharp2023}, with larger moduli being more robust to accumulated noise. We can think of the RNS representation as an $N \times \ell$ matrix, with $\ell$ being the number of $\textit{levels}$ or $\textit{towers}$ in the current ciphertext. 
\par CKKS supports several arithmetic operations like addition and multiplications between plaintext polynomials or ciphertexts. Additions and multiplications are simple point-wise operations between coefficients and therefore require the same number of levels in both operands. CKKS also supports $\textit{rotations}$ that cyclically rotate elements within the vector message by a specified amount, $r$. Notably, indexing vector elements is not efficiently supported and therefore ciphertext rotations are the primary way of computing fully connected layers and convolutions in neural networks.

\par Rotations and multiplications transform the ciphertext so that it cannot be decrypted by the original secret key, $s$. As a result, an auxiliary process known as ``key-switching'' is required, which is described in detail in the following section. This operation involves many NTTs, which are analogous to FFTs. A naive key-switching implementation can have poor operational intensity and bottleneck many practical applications. For example, recent work shows nearly 70\% of execution time is spent performing key-switching operations for ResNet-20~\cite{pmlr-v162-lee22e}.

\section{Hybrid Key-Switching}
\label{sect:HKS}
In this section, we describe the \textit{Hybrid} key-switching (HKS) algorithm, a crucial step after performing ciphertext rotations and ciphertext-ciphertext multiplications. Both operations convert a ciphertext encrypted by a secret key, $s$, into a new ciphertext only decryptable by a new secret key, $s'$. To continue computation, the ciphertext must be returned to a form decryptable by the original key, $s$. This process, known as key-switching, involves multiplying the ciphertext  $[\![ \mathbf{m} ]\!]_{s'}$ with a special evaluation key, $\mathbf{evk}$, which re-encrypts it from $s'$ to $s$. However, this process introduces significant noise growth, which can be managed using techniques proposed by Fan~\cite{fan2012somewhat} and later by Han~\cite{han2020better}. These techniques involve performing the $\mathbf{evk}$ multiplication at a higher modulus, reducing noise growth but increasing HE computational complexity with complex NTTs and RNS basis extensions.




HKS \cite{han2020better} is a generalization of existing key-switching techniques that let the user trade off the complexity of key-switching with the size of the $\mathbf{evk}$. It has been widely adopted both in software implementations \cite{Mouchet:299025, cryptoeprint:2022/915} and in recent HE accelerators \cite{samardzic2022craterlake, kim2022bts, kim2022ark, 10070953} yet the aforementioned trade-off, especially in hardware, is still not well understood.


\subsection{Overview}

\begin{figure*}
\centering
\captionsetup{justification=centering}
  \begin{minipage}[b]{0.6\textwidth}
  \centering
    \includegraphics[width=\textwidth]{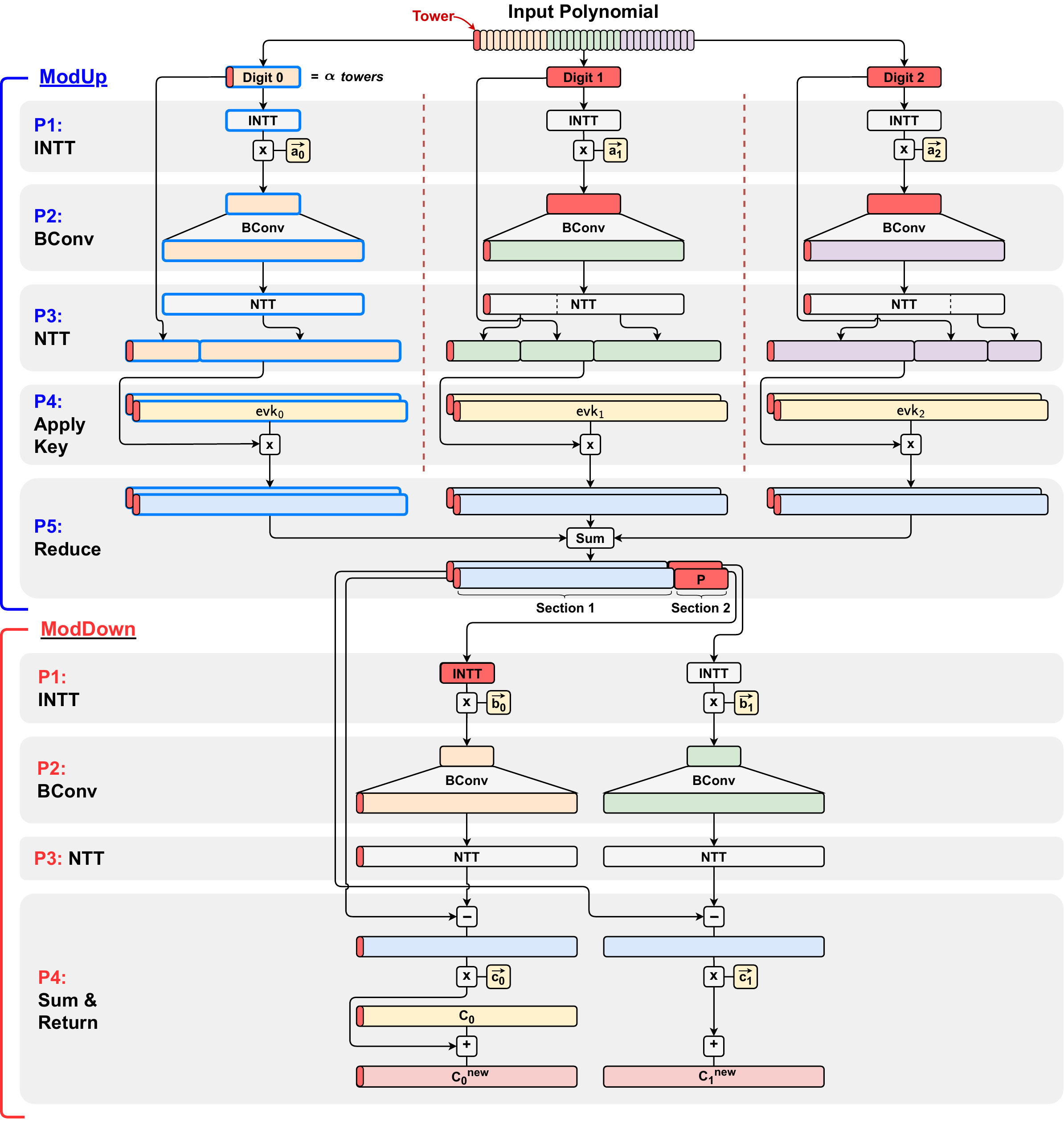}
    \caption{Hybrid key-switching dataflow diagram for parameters, $\ell=33, \ \alpha=11, \ dnum=3.$}
    \label{fig:main-hks}
  \end{minipage}\hspace{0.5em}%
  \begin{minipage}[b]{0.34\textwidth}
    \includegraphics[width=\textwidth]{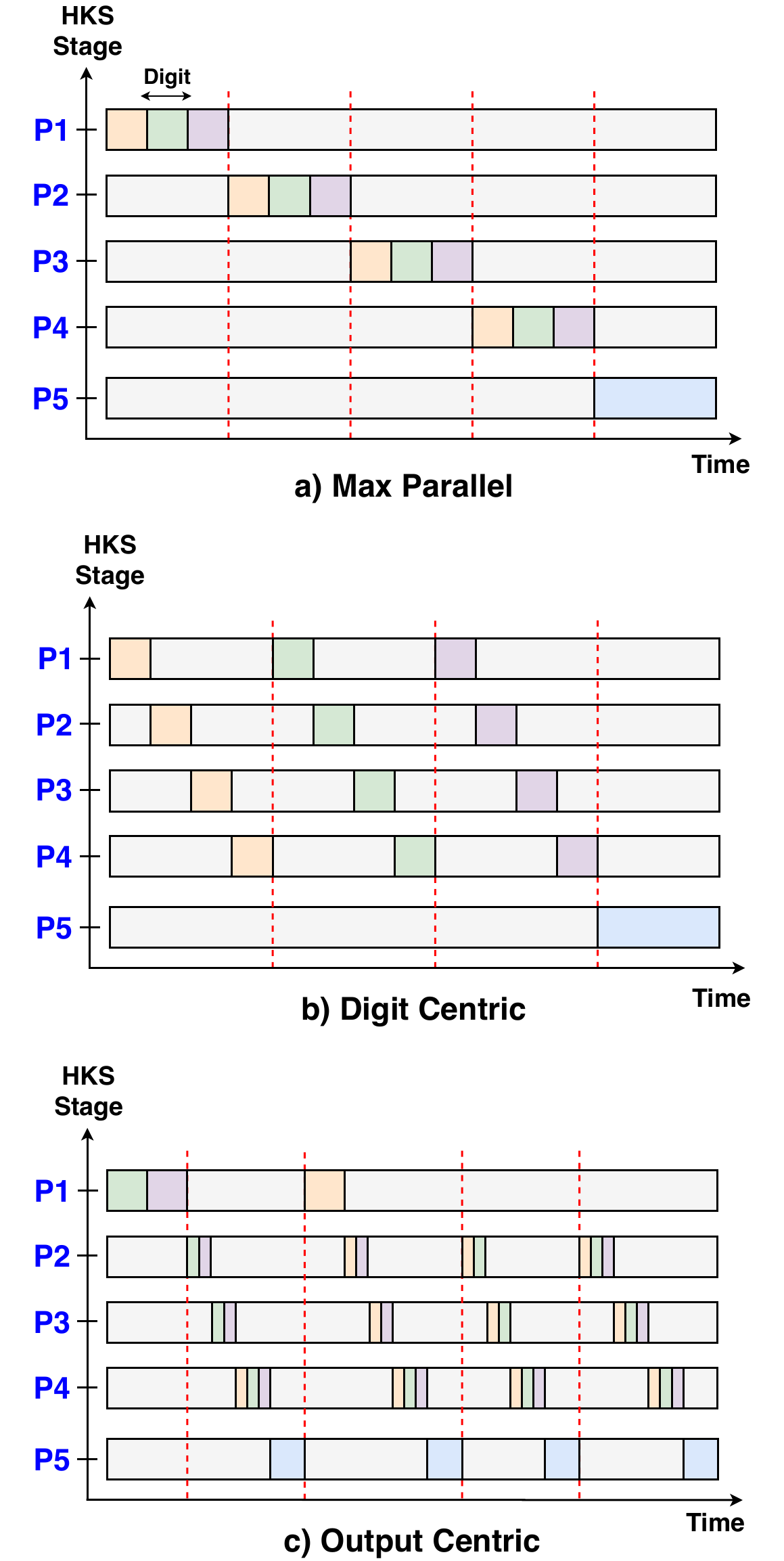}
    \caption{High-level $\textit{ModUp}$ timing diagrams for the three proposed dataflows.}
    \label{fig:main-hks-timing}
  \end{minipage}
  \vspace{-1.25em}
\end{figure*}

Broadly, HKS is composed of two phases that we denote as $\textit{ModUp}$ and $\textit{ModDown}$. The $\textit{ModUp}$ phase can be broken down into five consecutive stages, P1-P5, and the $\textit{ModDown}$ phase into four, P1-P4. 
We use terms such as $\textit{ModUp\//Down\_Pi}$ to refer to the $\textit{i'th}$ stage of the $\textit{ModUp\//Down}$ phase. 
These stages can be seen in Figure \ref{fig:main-hks}. We loosely adopt this terminology from Han~\cite{han2020better} to intuitively represent that the first phase of HKS is geared towards extending the RNS base from $Q_\ell$ to $P Q_\ell$, whereas the second phase is focused on reducing the modulus back from $P Q_\ell$ to $Q_\ell$. 

We consider an input polynomial as a matrix of size $(N \times \ell)$, which changes shape throughout the key-switching process, depending on the chosen parameters. Figure \ref{fig:main-hks} represents the HKS dataflow for a specific parameter set; the core stages of HKS remain the same across different parameter choices. 
In Section \ref{sect:benchmark}, we explore a wider range of parameter sets and note that the insights we gain from analyzing this specific case remain applicable. We highlight a single tower of the input polynomial in red in Figure \ref{fig:main-hks}, with the widths of subsequent stages accurately reflecting the change in sizes of the original $(N \times \ell)$ input polynomial. For example, a preliminary step in HKS decomposes the input polynomial into $dnum$ digits, each of size roughly, $\alpha = \lceil \sfrac{(L+1)}{dnum} \rceil$, reshaping the input matrix to $(N \times \ell)$. In Figure \ref{fig:main-hks}, for which $dnum=3$, the input polynomial is decomposed into $\textit{three}$ different colored digits, with each digit being $11$ towers wide. 



\subsection{$\textit{ModUp}$}

$\mathbf{(P1)} \hspace{0.2em} \textit{INTT}$: Following the initial digit decomposition step, each tower must undergo an $\textit{INTT}$, which converts the polynomial from the \textit{evaluation} domain to the \textit{coefficient} domain. This conversion has a complexity of $\mathcal{O}(N \log N)$, similar to the FFT, and is applied separately to all $\ell$ towers. 

$\mathbf{(P2)} \hspace{0.2em} \textit{BConv}$: Each digit, now in the coefficient domain, goes through a basis conversion to change the set of primes representing a polynomial. The number of towers is extended from $\alpha$ to $\beta$, where $\beta = \ell + K - \alpha$, and K is the number of towers in moduli $P$.
The number of modular multiplications in this stage is roughly $N \times \alpha \times \beta$ for each of the $dnum$ digits.

$\mathbf{(P3)} \hspace{0.2em} \textit{NTT}$: After basis extension, the digits once again return to the evaluation domain through the $\textit{NTT}$. The computational complexity, similar to the $\textit{INTT}$ stage, is $\mathcal{O}(N \log N)$, performed independently for $\beta \times dnum$ towers. This output is concatenated with the original digit to form a new polynomial modulo $P Q_\ell$. Our new matrix is now of size $dnum \times N \times (\ell + K)$.

$\mathbf{(P4)} \hspace{0.2em} \textit{Apply Key}$: Now that the polynomial is in a larger modulus, we can multiply it with $\mathbf{evk}$ with negligible added error. This is a point-wise operation with the $\mathbf{evk}$ being of shape $dnum \times 2 \times N \times (\ell + K)$. In practice, key sizes typically range from $100$MB to $400$MB (see Table \ref{tab:design_point}).

$\mathbf{(P5)} \hspace{0.2em} \textit{Reduce}$: The last stage in the $\textit{ModUp}$ phase sums each digit's output from  ${P4}$
into one final matrix of size $2N(\ell+K)$.

\subsection{$\textit{ModDown}$}

We now have two polynomials, each of size $N \times (\ell + K)$ that must be reduced to their original size of $N \times \ell$ for further computation. This process begins by taking the last $K$ towers of each polynomial and performing a similar series of $\textit{INTT} \rightarrow \textit{BConv} \rightarrow NTT$ operations. Here, each $\textit{BConv}$ operation converts the number of towers from $K$ to $\ell$. The runtime complexity of these three stages is $2 K \times N \log N$, $2 N \times K \times \ell$, and $2 \ell \times N \log N$, respectively. These polynomials go through one final scalar-tower multiplication and summation to finish out the key-switching process.

\section{CiFlow: A Taxonomy and Analysis \\of HKS Dataflow}
\label{sec:method}
We propose three dataflows for the HKS algorithm: Max-Parallel (MP), Digit-Centric (DC), and Output-Centric (OC). These dataflows differ in their sequence of instructions, reuse of loaded and computed data, intermediate data generation, and off-chip memory interaction. Assuming unlimited on-chip memory, the performance gap between these dataflows would decrease significantly. This is because all inputs and intermediate data could be stored on-chip, making MP even more advantageous than DC and OC due to its highly parallel nature. (E.g., this is the dataflow Cheetah~\cite{reagen2021cheetah} used to demonstrate the FHE overhead could be overcome with large chips.) Our goal is to show that by changing the sequence of operations we can have the same performance using a smaller on-chip memory and less off-chip bandwidth. In this section, we provide a detailed explanation of the principles and strategies behind each dataflow. Later, we evaluate their performance by varying bandwidth and computational throughput. Our analysis highlights how dataflow optimizations help to save memory and bandwidth resources.

\subsection{Max-Parallel Dataflow (MP)}
This dataflow is designed to prioritize kernel parallelism at all costs. Each tower is loaded onto the on-chip memory for a single operation, and each operation is executed on all input towers sequentially before processing the next step. However, this approach comes with a drawback. Specifically, in \textit{ModUp\_P2} and \textit{ModDown\_P2}, the generated intermediate data after BConv becomes extremely large, even though only one of the BConv output towers is required for calculating each output tower. For BTS3, at least 675MB of on-chip memory is required to prevent excessive load and stores to the off-chip.
MP was used by prior work (e.g., Cheetah \cite{reagen2021cheetah}, and HEAX \cite{riazi2020heax}), and we use it as the baseline HKS implementation.

\subsection{Digit-Centric Dataflow (DC)}
This dataflow adopts a ``one-digit-at-a-time'' approach, where each digit is loaded onto the chip and all possible calculations involving that digit are performed before moving on to the next digit. As illustrated in Figure \ref{fig:main-hks-timing}(b), all P1 to P5 calculations for a single digit are done in sequence, maximizing data reuse. The blue frames in Figure \ref{fig:main-hks} show the intermediate data generated from processing a single digit in the \textit{ModUp} section, which is used for calculating a partial product of the \textit{ModUp\_P5} step. As depicted in Figure \ref{fig:main-hks}, BConv still expands the digit from $\alpha$ to $\beta$ towers, generating a partial product of the output that can either be stored on-chip for later reduction to save bandwidth or sent off-chip to minimize on-chip memory requirements. Once \textit{ModUp} is completed, \textit{ModDown} follows the same approach.
For BTS3, which is the largest benchmark, DC requires 255MB of on-chip memory, which is 62\% less compared to MP dataflow.
\vspace{-0.1em}
\subsection{Output-Centric Dataflow (OC)}
Finally, we propose Output-Centric (OC) dataflow, which is optimized for data movement and on-chip storage, leveraging insights from the first two solutions. OC focuses on computing one output tower at a time, efficiently utilizing on-chip resources. In this dataflow, the \textit{ModUp} stage has two sections: Section1 computes \textit{ModUp} output towers in modulo $Q$, with $dnum-1$ digits going through \textit{ModUp\_P2 \& ModUp\_P3} step and one tower from the last digit being bypassed through \textit{ModUp\_P2 \& ModUp\_P3}. As shown by the red towers in Figure~\ref{fig:main-hks}, to generate the first output tower, the first digit is bypassed, while the other two digits must pass through the \textit{ModUp\_P2} stage. Section2 computes \textit{ModUp} output towers in mod $P$, requiring all digits to pass through \textit{ModUp\_P2 \& ModUp\_P3} for a single output calculation. 

\par
In both sections, the computation is optimized to minimize on-chip memory requirements. Since only one output tower is calculated at a time, the entire computation of \textit{ModUp\_P2 \& ModUp\_P4} is unnecessary. The red towers in Figure~\ref{fig:main-hks} represent the computations needed to generate one output tower, with \textit{ModUp} being in Section1. As shown, there is no need to do all calculations of \textit{ModUp\_P2 \& ModUp\_P4}, minimizing the memory requirement and off-chip data communication. Additionally, for \textit{ModUp\_P5}, only one partial tower is calculated at a time, allowing them to be on-chip for accumulation and only store back the accumulation result.


In Section2, where all digits are required for calculating a single output tower in \textit{ModUp}, we have used the following strategy to manage the on-chip memory limitation. Since the INTT of the first $dnum-1$ digits are already on-chip, we compute the partial sum with those digits and then the final digit is loaded to compute the last partial sum and the final output towers.
This approach reduces off-chip memory interaction and on-chip memory requirement.
\par

\textit{ModDown} has the same approach as \textit{ModUp}, loading all towers related to $[\mathtt{P}]_{\mathcal{C}}$ on-chip. Calculating one output tower at a time eliminates the expansion of \textit{ModDown\_P2}, enhancing the efficiency of HKS with a small on-chip memory.

\begin{table}[t!]
\captionsetup{justification=centering}
  \caption{DRAM transfers (MB), including $\mathbf{evk}$ with 32MB on-chip memory and Arithmetic Intensity (AI) in ops/byte. 
  }
   \small
   \centering
\begin{tabular}{l|cc|cc|cc}
\toprule\toprule
\multicolumn{1}{c|}{\multirow{2}{*}{\textbf{Benchmark}}} & \multicolumn{2}{c|}{\textbf{MP}} & \multicolumn{2}{c|}{\textbf{DC}} & \multicolumn{2}{c}{\textbf{OC}} \\
\multicolumn{1}{c|}{}                                    & MB             & AI              & MB             & AI              & MB            & AI              \\
  \midrule
BTS1                                                     & 600          & 1.81           & 600          & 1.81           & 420           & 2.59           \\
BTS2                                                     & 1352            & 1.14           & 1278            & 1.2           & 716           & 2.15           \\
BTS3       & 1850               & 1.00           & 1766           & 1.04           & 1119 & 1.65           \\
ARK                                                      & 432            & 1.05           & 356            & 1.27           & 180            & 2.52           \\
DPRIVE                                                   & 365            & 1.26           & 336            & 1.37           &170            & 2.71     \\      
   \bottomrule
\end{tabular}
   \label{tab:load_store}
   \vspace{-1em}
\end{table}

\subsection{Dataflow Comparison: Arithmetic Intensity}
Table~\ref{tab:load_store} evaluates the off-chip data movement,
including $\mathbf{evks}$ and input/output data, of each benchmark. 
Here, the assumed on-chip memory is 32MB with $\mathbf{evks}$ being streamed on-chip.
Across all benchmarks, we see that OC can significantly reduce the total off-chip traffic.
The number of operations per HKS benchmark is independent of dataflow.
Thus, we can see that the arithmetic intensity (AI) of each is also significantly improved with OC. 

In a recent study, MAD~\cite{mad} introduced techniques to optimize and accelerate FHE. Analyzing their optimizations, we find their solution is analogous to our proposed DC dataflow. According to the reported arithmetic intensity in Table~\ref{tab:load_store}, leveraging the OC dataflow results in 1.43$\times$ to 2.4$\times$ more arithmetic intensity than MP and 1.43$\times$ to 1.98$\times$ more than DC. Furthermore, MAD~\cite{mad} (and other FHE accelerators) also consider a technique for key compression, which halves the off-chip data movement of keys. Incorporating key compression in our approach will further boost our AI to 3.82.

\section{Methodology}
This section provides an overview of our dataflow implementations. We will start by describing the hardware architecture and then explain the benchmarks used for performance evaluation. Finally, we will introduce the software framework employed for generating HKS instructions for different benchmarks and dataflows.

\subsection{Hardware Simulation}
\label{sec:rpu}
\begin{figure}[t]
  \centering
  \includegraphics[width=0.98\linewidth, scale=1]{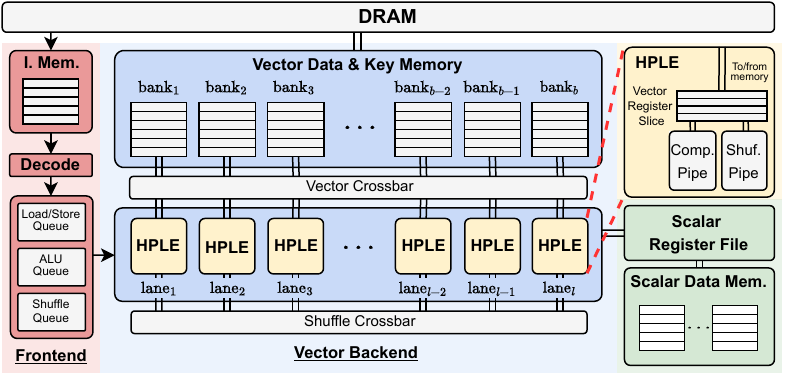}
  \caption{Microarchitecture of the RPU.}
  \label{fig:rpu}
  \vspace{-1em}
\end{figure}

We implement the dataflows introduced in Section \ref{sec:method} on the RPU \cite{soni2023rpu} and conduct an extensive evaluation in Section \ref{sec:eval}. The RPU is a vector processor supporting Ring-Learning-With-Errors (RLWE)-based algorithms, including HE, with 64, 64-bit vector registers, 64, 64-bit scalar registers, a 32MB vector data memory, and a 1MB scalar data memory, operating at 1.7GHz. These exist alongside an additional modulus register file, which stores a set of RNS moduli. Figure \ref{fig:rpu} presents the RPU microarchitecture. It consists of a front-end to handle instruction fetching, decoding, and control logic, and a backend that contains the high-performance large arithmetic word engines (HPLEs) to efficiently perform HE operations. A parameterized simulator is used to test different RPU configurations (e.g., SRAM capacity and computational throughput), and supports arbitrarily large parameter sets.

This work uses a $2\times$ smaller RNS moduli compared to the original RPU paper~\cite{soni2023rpu}. 
Here, we assume 128 HPLEs and have modified the RPU's associated instruction set architecture (ISA), $\mathtt{B512}$, to support a vector length of $1K$, referred to as $\mathtt{B1K}$, to maintain high throughput and keep compute units occupied. Longer vectors make hardware efficient, e.g., taking pressure off the frontend and improving compute utilization. $\mathtt{B1K}$ consists of $28$ instructions ranging from general purpose point-wise arithmetic operations to HE-specific \textit{shuffle} instructions for (i)NTT kernels. 


Since FHE is data-oblivious, all memory addresses are known at compile time, and the behavior is independent of input values. This property enables data prefetching through decoupling, where the compiler can order memory requests to overlap data movement with computation. This optimization leverages the deeply decoupled microarchitecture of the RPU, which employs three distinct queues to fetch independent compute, shuffle, and memory instructions in parallel. This overlap helps mask latency and improve overall performance.



\subsection{Benchmark Selection}
\label{sect:benchmark}
We use the RNS variant of HKS, as presented in Han~\cite{han2020better}, which optimizes the key-switching implementation by leveraging RNS decomposition for enhanced efficiency. The security of HE is defined by $\lambda$, which is a function of \(\frac{N}{logQP}\). As mentioned in Cheon~\cite{cheon2019hybrid}, 128-bit security is required to defend against dual attacks and to be able to extract the real data from the encrypted output. BTS~\cite{kim2022bts} simulates the multiplication time per slot by sweeping $N$, $L$, and $\lambda$ to identify the optimal configurations for 128-bit security, and refers to these points as BTS1, BTS2, and BTS3. ARK \cite{kim2022ark} uses a smaller polynomial degree, $2^{16}$, while increasing the number of digits to 4 to satisfy the 128-bit security. As in recent studies \cite{kim2022bts, kim2022ark, bossuat2021efficient, 10070953} we provide 128-bit security using proper FHE parameterization from BTS\cite{kim2022bts}, ARK\cite{kim2022ark}, and DPRIVE~\cite{baa}, to evaluate the performance of our dataflows. Table \ref{tab:design_point} summarizes the parameters of the benchmarks used in our research. 
\begin{table}[t]
   \caption{Parameters satisfying 128-bit security.} 
   \label{tab:design_point}
   \small
   \centering
   \begin{tabular}{lccccccc}
   \toprule\toprule
   \textbf{Benchmark}~   & \textbf{N}            & \textbf{$k_l$} & \textbf{$k_p$} & \textbf{dnum} & \textbf{$\alpha$} & \textbf{evk Size} & \textbf{Temp data}\\ 
   \midrule
   \textbf{BTS1}   & 2\textsuperscript{17} & 28                & 28                & 1             & 28                             & 112MB       &196MB      \\
\textbf{BTS2}   & 2\textsuperscript{17} & 40                & 20                & 2             & 20                             & 240MB         &400MB    \\
\textbf{BTS3}   & 2\textsuperscript{17} & 45                & 15                & 3             & 15                             & 360MB     &  585MB      \\
\textbf{ARK}    & 2\textsuperscript{16} & 24                & 6                 & 4             & 6                              & 120MB   &192MB          \\
\textbf{DPRIVE} & 2\textsuperscript{16} & 26                & 7                 & 3             & 9     & 99MB  &163MB   \\
   \bottomrule
   \end{tabular}
   
\vspace{-4mm}
\end{table}

\subsection{Software Framework}
To generate HKS code for different benchmarks and dataflows, we break down the HKS algorithm into steps highlighted in Figure \ref{fig:main-hks} and generate instructions for each step independently, based on the $\mathtt{B1K}$ ISA. Our simulation framework includes two distinct tasks: memory and compute. Memory tasks handle data transfer between off-chip and on-chip, while compute tasks correspond to steps in the HKS algorithm, executed on the RPU~\cite{soni2023rpu}. The challenge of our approach lies in effectively managing task dependencies and the sequence of instructions. Each task may rely on one or more compute and memory tasks, which must complete their execution before the dependent task can proceed. These dependencies stem from a variety of reasons, such as the need to fetch data from off-chip memory or to create space for subsequent operations using a memory task. Moreover, some tasks depend on data generated by a compute task as part of their input. These dependencies may vary based on the on-chip memory, the specific dataflow, and the benchmarks used. Using the software framework, we generate instructions for each configuration and dataflow and define their dependencies. The framework has two distinct queues, one for memory tasks and one for compute tasks. The tasks at the front of each queue are fetched and executed in parallel once all the task's dependencies are resolved. If there are no dependencies between a memory task and a compute task, the off-chip data movement can be masked by the computation on the RPU.

\subsection{Dataflow Simulation}
\label{sect:dataflow}
We have implemented and analyzed all given benchmarks in Table \ref{tab:design_point} for each dataflow described in Section \ref{sec:method}. In Section \ref{sec:method}, we noted that with unlimited on-chip memory, the performance of the dataflows tends to be nearly identical. However, with limited on-chip memory, this would no longer be true because the inputs and intermediate data cannot all fit on the chip at once. This makes the sequence of operations and the data movement between on and off-chip memory crucial. When dealing with small on-chip memory and increased intermediate data, the performance of DC may become close to MP due to more interactions with off-chip memory.

\par
Our goal is to reduce interactions with off-chip memory by increasing on-chip data reuse. In general, when sufficient on-chip memory is available, INTT outputs can be stored on-chip and reused. However, with limited on-chip storage, the INTT output must be stored off-chip and reloaded for subsequent computations, increasing the off-chip data movement. For example, in BTS3, using the MP implementation, INTT is applied to all 45 input towers, leaving no space for storing all INTT outputs on-chip. In contrast, using OC, INTT is applied to 30 towers, allowing the INTT outputs to be stored on-chip and used for later computation. 


In \textit{ModUp\_P5}, if on-chip space is sufficient, we prioritize storing towers related to $[P_0]_\mathcal{B}$ and $[P_1]_\mathcal{B}$, for optimized on-chip memory use and reduced off-chip memory access during subsequent computations. Additionally, \emph{ModDown\_P2} and \emph{ModDown\_P3} will begin computations from towers that are already on-chip.  


\par

 
\par

\par


\section{Evaluation}
\label{sec:eval}

\begin{figure*}[t!]
  \centering
  \includegraphics[width=0.99\linewidth, scale=0.5]{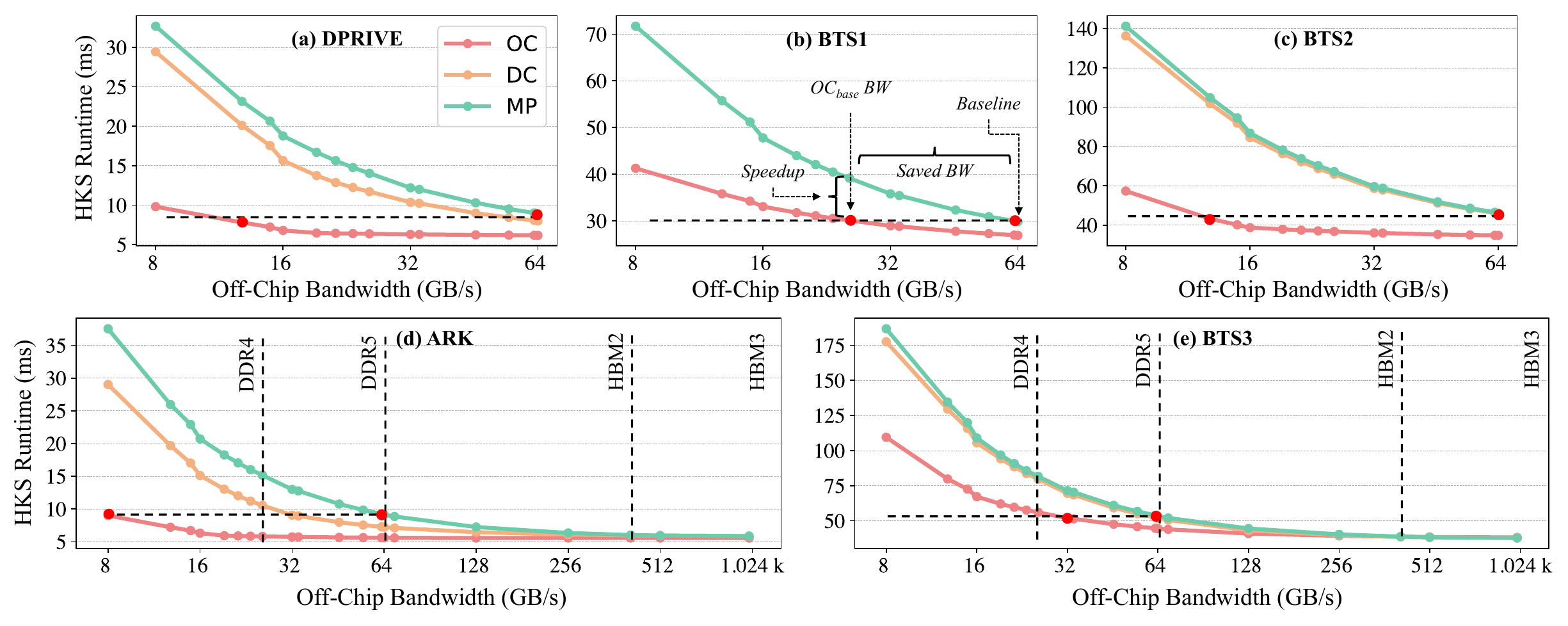}
  \vspace{-2mm}
  \caption{Quantifying latency reduction for MP, DC, and OC by increasing DRAM bandwidth for the five given benchmarks.}
  \label{fig:hks_latency}
  \vspace{-5mm}
\end{figure*}

In this section, we evaluate HKS performance following the three dataflows described above, using five benchmarks listed in Table~\ref{tab:design_point} from recent work (BTS~\cite{kim2022bts}, ARK~\cite{kim2022ark}, and DPRIVE~\cite{baa}).
The evaluation is done using the RPU~\cite{soni2023rpu} as the target hardware platform of our implementations, including the modifications mentioned in Section~\ref{sec:rpu}. While the performance results are specific to the RPU, the general insights and takeaways from the optimizations are broadly applicable.
As mentioned earlier, one of the challenges in HKS is properly handling large amounts of input, intermediate data, and $\mathbf{evks}$.
In fact, prior work \cite{kim2022bts, kim2022ark, jung2021over} has found that HE is memory bound.
We re-examine this performance limitation considering the different proposed dataflows.
First, we assume a large 392MB on-chip SRAM, with 32MB allocated for data and the rest for $\mathbf{evks}$, pre-loaded on-chip.
We sweep bandwidth across all three dataflows to demonstrate the bandwidth saving achievable with OC while maintaining performance equal to MP and DC at higher bandwidths.
Next, we eliminate the on-chip SRAM dedicated to storing $\mathbf{evks}$, instead streaming them on-chip, as they are used only once per HKS, and saving substantial on-chip SRAM (up to $12.25\times$). We find that a small on-chip memory significantly degrades the performance of a naive implementation (MP), but that our optimizations (OC) result in minimal slowdown.
Finally, we increase the computational throughput of the RPU to understand how our proposed dataflow balances the communication-to-compute ratio.



\subsection{Dataflow Comparison: Saving Bandwidth}
Figure~\ref{fig:hks_latency} compares different HKS implementation runtimes across different dataflows for all benchmarks. 
Benchmarks are categorized into three groups:
ARK and DPRIVE (small, $2^{16}$ polynomial degree); 
BTS1 (single digit, lacks \textit{ModUp\_Reduce}), and
BTS2 and BTS3 (large, $2^{17}$ with three and four digits). 
We sweep the off-chip bandwidth from 8GB/s to 64GB/s, extending to 1TB/s for ARK and BTS3, representing the smallest and largest benchmarks, respectively.
This bandwidth range includes DDR4 (8GB/s to 25.6GB/s), DDR5 (32GB/s to 64GB/s), HBM2 (64GB/s to 410GB/s), and HBM3 (up to 1TB/s). 
In all benchmarks, the performance benefit of OC is large at low bandwidth and decreases as bandwidth increases. 
This is the result of RPU becoming compute bound, reducing the impact of off-chip communication.  
With OC, we show that by optimizing the dataflow we can achieve higher performance with less bandwidth.

Below we will delve into a detailed comparison between OC, MP, and DC runtimes, taking the MP implementation with 64GB/s off-chip bandwidth and on-chip pre-loaded $\mathbf{evks}$ as our \textit{baseline}. We will explore the bandwidth at which OC matches \textit{baseline} performance, denoting it as $OC_{base}$. 64GB/s was chosen as our reference point, given that it represents the peak DDR5 bandwidth in our evaluations, and highlighting cost-effective designs before moving to 
expensive memory technologies.


\subsubsection{DPRIVE \& ARK}
Figure~\ref{fig:hks_latency}(a) and (d), demonstrate a significant improvement in HKS runtime with OC dataflow.
The horizontal line in Figure~\ref{fig:hks_latency}(a) shows that OC operating at 12.8GB/s bandwidth matches the baseline, resulting in a 5$\times$ bandwidth saving. At this specific data point, OC is 2.57$\times$ and 2.96$\times$ faster than DC and MP, respectively. Before becoming compute bound, the RPU is idle for part of its execution time, waiting for dependent memory tasks to be completed. With 12.8GB/s off-chip bandwidth, OC causes the RPU to be idle for 20.87\% of its execution time, outperforming DC and MP with 68.62\% and 72.76\% idle times, respectively. This highlights OC's advantages when off-chip bandwidth is a concern and its efficiency in minimizing idle time. 

In the case of ARK, the OC implementation with 8GB/s off-chip bandwidth achieves the same performance as MP and DC with 64GB/s bandwidth, resulting in 8$\times$ and 5$\times$ bandwidth savings, respectively. With 8GB/s off-chip bandwidth, OC outperforms MP and DC runtime by notable factors of 4.16$\times$ and 3.22$\times$ and being 2.25$\times$ less idle cycle than MP.




\subsubsection{BTS1}
As mentioned above, for BTS1 with one digit, MP and DC have the same implementation. In Figure~\ref{fig:hks_latency}(b), OC matches the baseline performance by 25.6GB/s off-chip bandwidth, saving 2.5$\times$ bandwidth. At this point, OC is 1.3$\times$ faster than MP and 2.1$\times$ less idle.

\subsubsection{BTS2 \& BTS3} By analyzing BTS2 and BTS3, we observe that as the benchmark size increases, DC and MP converge due to the large data size during $\textit{BConv}$ expansion and more polynomial additions for reduction.


According to Figure~\ref{fig:hks_latency}(c), BTS2's runtime with OC dataflow and less than 12.8GB/s off-chip bandwidth matches the baseline, saving 5$\times$ bandwidth. With 12.8GB/s bandwidth, OC is 2.35$\times$ faster than DC and 2.42$\times$ faster than MP.


For BTS3, the performance gap between OC and the other two dataflows decreases as the RPU becomes more compute bound.
Despite the decreasing gap between OC and MP/DC, due to the increased input and intermediate data in BTS3, OC still outperforms MP/DC. With 32GB/s bandwidth, OC is 1.3$\times$ faster and 1.7$\times$ less idle than MP and DC, while matching the baseline's performance, resulting in a 2$\times$ bandwidth saving.

\subsubsection{Comparison Across Benchmarks}

\begin{table}[t!]
   \caption{OC bandwidth ($OC_{base}$; in GB/s) required to achieve the same performance as MP using 64GB/s BW.
   Speedup reports the performance improvement OC achieves over MP at the listed bandwidth ($OC_{base}$).}
   \small
   \centering
   \begin{tabular}{l|ccccc}
   \toprule\toprule
   \textbf{Benchmark}~   & \textbf{$\mathbf{OC_{base}}$} & \textbf{Saved} & \textbf{OC} & \textbf{MP} & \textbf{OC} \\ 

   \textbf & (GB/s) & \textbf{BW} & (ms)& (ms)& \textbf{Speedup}  \\

   \midrule
   \textbf{BTS1}   & 25.60  & 2.50x & 30.08 & 39.13& 1.30x    \\
\textbf{BTS2}  & 12.80& 5.00x& 43.24&104.85&2.42x \\
\textbf{BTS3}   & 32.00&2.00x&51.87&71.50&1.37x\\
\textbf{ARK}   &8.00&8.00x& 9.01&37.54&4.16x \\
\textbf{DPRIVE} &12.80&5.00x& 7.81&23.15&2.96x  \\
   \bottomrule
   \end{tabular}
   \label{tab:concl}
   \vspace{-4mm}
\end{table}

Table~\ref{tab:concl} summarizes the previous discussion, with ``\emph{$OC_{base}$}'' indicating the bandwidth where OC matches the \textit{baseline} performance (MP with 64MB/s bandwidth), and ``\emph{Saved BW}'' showing the bandwidth savings achieved by OC. Runtimes are reported based on the \emph{$OC_{base}$} bandwidth. 
Notably, BTS1 has less speedup due to its larger size compared to ARK and DPRIVE and more data movement for the reduction than BTS2 and BTS3. This can also be seen in Table~\ref{tab:load_store}, where the AI (Arithmetic Intensity) improvement for BTS1 in OC is less than other benchmarks.
\par
Many recent HE algorithmic optimizations, e.g., RNS-based HKS \cite{han2020better}, which is used in this paper, require a higher bandwidth. Therefore, reducing off-chip bandwidth is crucial. 
The OC dataflow results in 1.30$\times$ to 4.16$\times$ speedup over a simple MP implementation and saves 2$\times$ to 8$\times$ off-chip bandwidth. 

\vspace{-0.5mm}

\subsection{Streaming Evaluation Keys: Trading SRAM for Bandwidth}
Analyzing the computations of HKS, the input data and intermediate data ($\textit{BConv}$ output) are reused for calculating multiple output towers, making it advantageous to keep them on-chip to reduce off-chip communication. 
On the other hand, the \textbf{evks} are large, ranging from 99MB (DPRIVE) to 360MB (BTS3), and are only used once per HKS.
Consequently, it can be more practical and area efficient to store these large $\mathbf{evks}$ off-chip and load them on-chip in a streaming fashion. 
To study the effects of streaming $\mathbf{evks}$, we reserve a fraction of off-chip bandwidth and dedicate it to loading the $\mathbf{evks}$ on-chip. In this case, we only have 32MB of on-chip memory to store data and capture on-chip reuse.
To determine the required bandwidth allocation for $\mathbf{evks}$, we calculate the ratio of the number of $\mathbf{evks}$ to the number of loaded/stored data. 

\begin{figure}[t!]
  \centering
  \includegraphics[width=0.99\linewidth, scale=0.5]{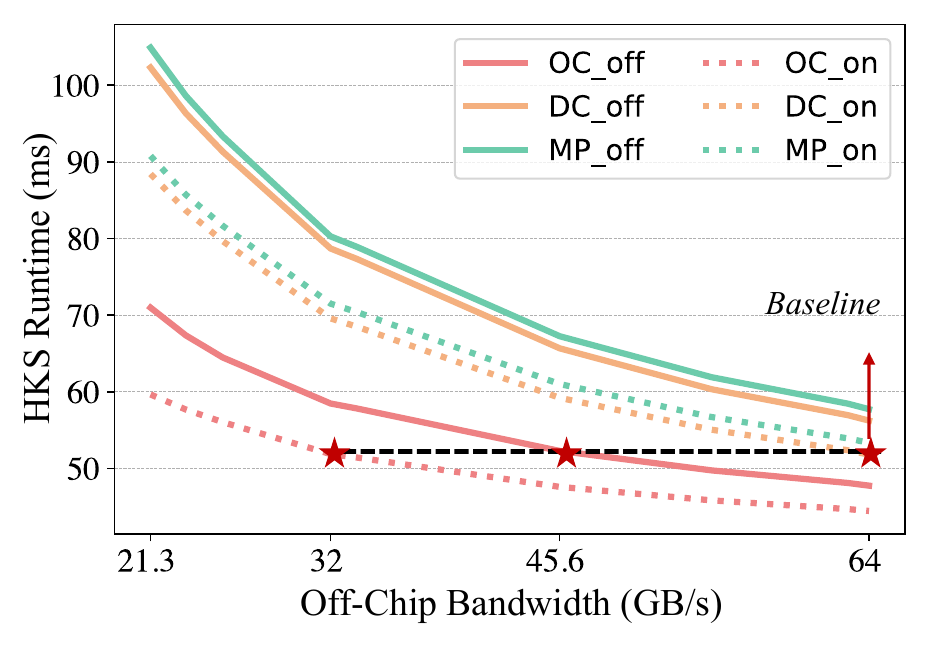}
  \vspace{-1.75em}
  \caption{HKS runtime for BTS3 with $\mathbf{evks}$ being off-chip.}
  \label{fig:bts3_key}
  \vspace{-1em}
\end{figure}

\begin{figure}[t!]
  \centering
  \includegraphics[width=0.99\linewidth, scale=0.5]{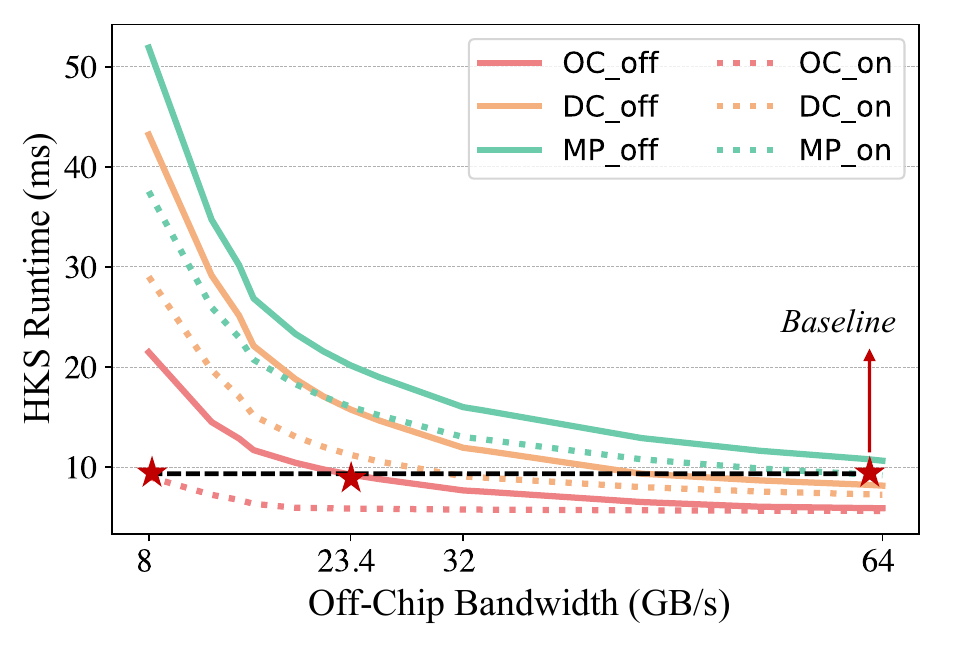}
  \vspace{-1.75em}
  \caption{HKS runtime for ARK with $\mathbf{evks}$ being off-chip.}
  \label{fig:Ark_key}
  \vspace{-1.75em}
\end{figure}

Figures~\ref{fig:bts3_key} and ~\ref{fig:Ark_key} illustrate the effects of storing $\mathbf{evks}$ off-chip and HKS runtime as a function of bandwidth. The dotted lines represent the runtime with pre-loaded $\mathbf{evks}$ stored on-chip. As shown in the figures, storing $\mathbf{evks}$ off-chip maintains the same trend but with shifted absolute values due to the increased bandwidth pressure from streaming $\mathbf{evks}$. According to Table \ref{tab:concl}, for BTS3, 32GB/s is the \textit{$OC_{base}$} bandwidth (equal performance as baseline). However, Figure~\ref{fig:bts3_key} demonstrates that by streaming $\mathbf{evks}$ on-chip, the OC dataflow can achieve baseline performance with a higher bandwidth of 45.62GB/s. Similarly, Figure \ref{fig:Ark_key} shows this concept for ARK, where a bandwidth of 23.4GB/s allows ARK to match baseline performance. While in both these cases more bandwidth is required to match performance, we argue the increase is minor compared to the $12.25\times$ reduction in on-chip SRAM.

\begin{figure}[t]
  \centering
  \includegraphics[width=0.99\linewidth, scale=0.5]{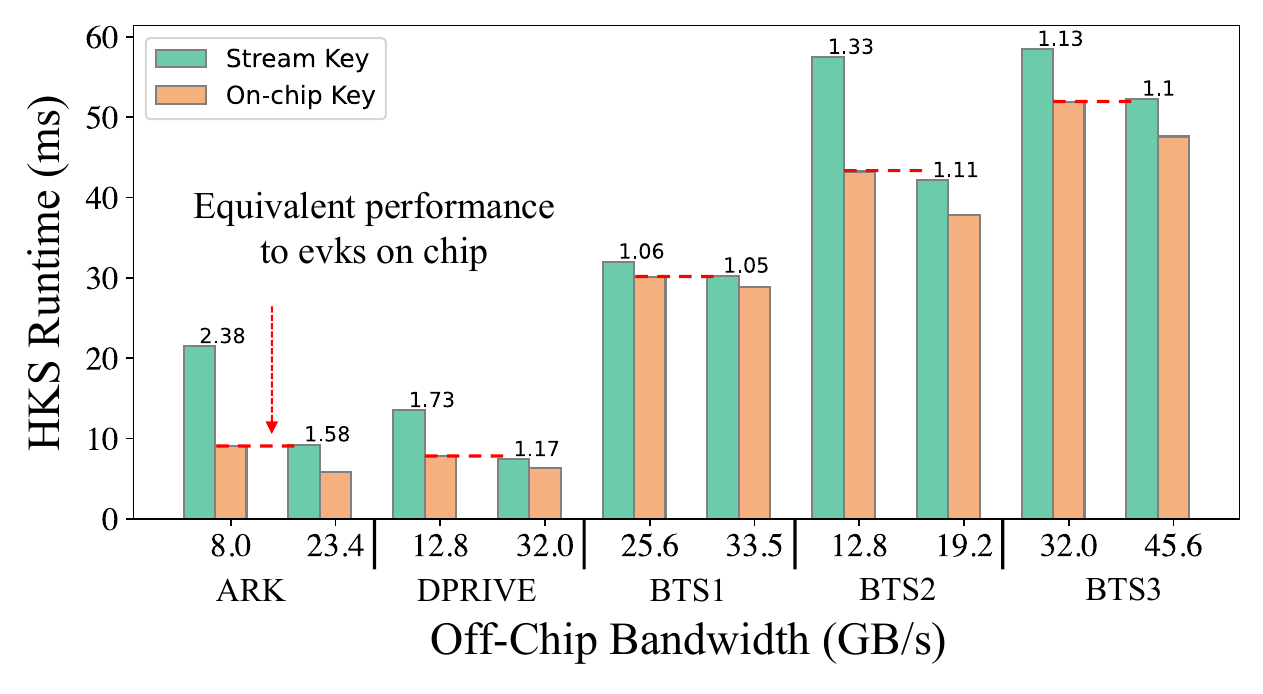}
  \vspace{-1.75em}
  \caption{HKS runtime for OC with $\mathbf{evks}$ streamed from off-chip vs. $\mathbf{evks}$ stored on-chip, and the equivalent bandwidth.}
  \label{fig:keystream}
  \vspace{-1.5em}
\end{figure}

Figure~\ref{fig:keystream} shows the slowdown for the OC dataflow of each benchmark when streaming $\mathbf{evks}$ from off-chip. The results are shown for two bandwidths per benchmark, as indicated by the clustered bars (e.g., 8 and 23.4 for ARK). 
The first bandwidth corresponds to the $OC_{base}$ bandwidth specified in Table~\ref{tab:concl}, where the performance is equal to the baseline, assuming $\mathbf{evks}$ to be on-chip. 
The second bandwidth indicates the required bandwidth to attain equivalent performance when streaming the $\mathbf{evks}$ from off-chip.
The slowdown for DPRIVE is less than ARK due to the smaller ratio of $\mathbf{evks}$ to the loaded/stored data (0.66 for ARK and 0.5 for DPRIVE). Among the large benchmarks (BTS1, BTS2, and BTS3), BTS2 has the most slowdown, 1.33$\times$, since the mentioned ratio is more than other benchmarks. As shown in Figure~\ref{fig:keystream}, with the OC dataflow, and 1.3$\times$ (BTS1) to 2.9$\times$ (ARK) more bandwidth we can achieve the same performance as storing the $\mathbf{evks}$ on-chip while saving 12.25$\times$ on-chip memory. However, compared to the original 64GB/s MP implementation with $\mathbf{evks}$ on-chip, the OC dataflow with keys off-chip still saves 1.4$\times$ up to 3.3$\times$ bandwidth for BTS3 and BTS2, respectively, to achieve the same performance.
In conclusion, storing $\mathbf{evks}$ off-chip saves 12.25$\times$ on-chip memory and by implementing OC dataflow, we can save up to 3.3$\times$ bandwidth to have the same performance as the baseline. Additionally, storing $\mathbf{evks}$ off-chip and keeping 32MB on-chip memory for data, decreases the RPU area from $401.85mm^2$ to $41.85mm^2$.

\subsection{Sensitivity Analysis: Balancing Bandwidth and Computational Throughput}
In the last section, we evaluated the three proposed dataflows and showed how data reuse enabled the OC optimization to significantly improve the runtime of HKS, saving considerable bandwidth even when streaming $\mathbf{evks}$. In this section we conduct a sensitivity study to investigate how the performance of HKS with OC is affected as we increase both the bandwidth, ranging up to 1TB/s (HBM3), and the computational throughput by up to 16$\times$ more. We evaluate HKS performance for ARK and BTS3 benchmarks, the smallest
and largest benchmarks, respectively. 
We refer to computational throughput as MODOPS (Modular Operations per Second).

\subsubsection{Bandwidth Analysis}
In Figure~\ref{fig:hks_latency}(d) and (e) we increase the off-chip bandwidth beyond 64GB/s (DDR5) up to 1TB/s (HBM3), with HPLEs fixed at 128, to study how scaling bandwidth affects HKS runtime for ARK and BTS3.
For both ARK and BTS3, the benefit from OC compared to the other two dataflows diminishes at bandwidths larger than 256GB/s. At this point, the off-chip data movement is mostly masked by computation on the RPU, limiting further performance gain from increasing bandwidth.

According to Figure~\ref{fig:hks_latency}(e), moving from 1TB/s to the \textit{$OC_{base}$} BW for BTS3, results in a 31.25$\times$ bandwidth saving, with only a 1.35$\times$ increase in runtime, while transitioning to the same bandwidth (32GB/s) with the MP dataflow leads to a 13.98$\times$ slower runtime than the 1TB/s implementation.

In Figure~\ref{fig:hks_latency}(d), we see that by employing OC dataflow for ARK, data movement becomes fully masked by computation after reaching a bandwidth of 128GB/s. With 8GB/s of off-chip bandwidth (\textit{$OC_{base}$}) and the OC implementation, we can save 16$\times$ bandwidth by being 1.6$\times$ slower compared to OC with 128GB/s. That is while for MP implementation, moving to 8GB/s causes a 5.17$\times$ increment in the runtime.


\subsubsection{Computational Throughput Analysis}
 \begin{figure}[t]
  \centering
  \includegraphics[width=1\linewidth, scale=0.8]{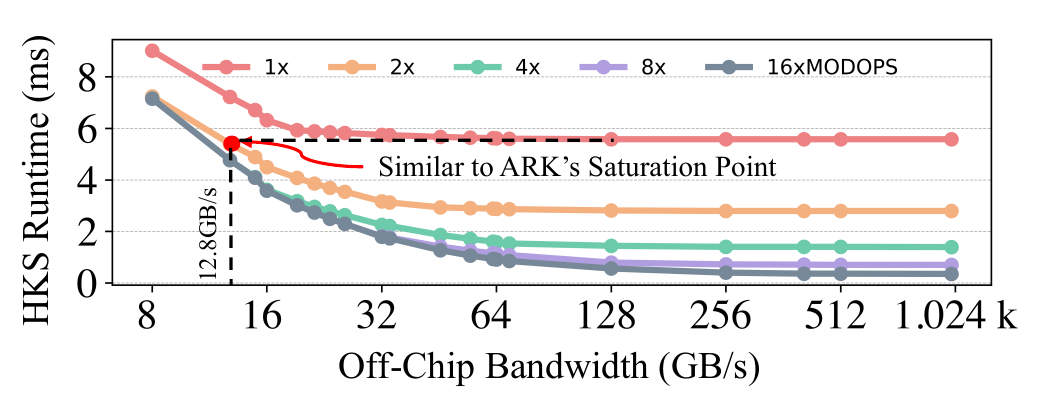}
  \vspace{-2em}
  \caption{Evaluating HKS runtimes for ARK using OC at different MODOPS, with evks on-chip.}
  \label{fig:arkfreqall}
\end{figure}

\begin{table}[t!]
\caption{OC, DC, and MP configuration for equivalent performance to ARK's saturation point.}

   \label{tab:ark_modops}
   \small
   \centering
\begin{tabular}{c|cccc}
 \toprule\toprule
\multicolumn{1}{c|}{\textbf{Dataflow}}                              & \textbf{\begin{tabular}[c]{@{}c@{}}BW\\ GB/s\end{tabular}} & \textbf{\begin{tabular}[c]{@{}c@{}}RPU\\ MODOPS\end{tabular}} & \textbf{\begin{tabular}[c]{@{}c@{}}Rel.\\ BW\end{tabular}} & \textbf{\begin{tabular}[c]{@{}c@{}}Rel.\\ MODOPS\end{tabular}} \\
   \midrule
\textbf{Sat. Point} & 128                                                        & 1.00x                                                         & 1.00x                                                      & 1.00x                       \\
   \midrule
\textbf{OC}                                                         & 12.80                                                      & 2.00x                                                         & 0.10x                                                      & 2.00x                                                              \\
\textbf{DC}                                                         & 54.64                                                      & 2.00x                                                         & 0.42x                                                      & 2.00x                                                             \\
\textbf{MP}                                                         & 128.00                                                     & 2.00x                                                         & 1.00x                                                      & 2.00x                                                       \\             
   \bottomrule
   \end{tabular}
   \vspace{-1em}
\end{table}
The RPU design includes 128 lanes, meaning 128 modular multipliers, which is 128$\times$ less functional units compared to BTS~\cite{kim2022bts}. To enhance the HKS runtime, which benefits from parallel computation where the calculation of each output tower is entirely independent of others, we simulate an accelerator with greater computational throughput by increasing the RPU's MODOPS (2$\times$, 4$\times$, 8$\times$, and 16$\times$). 
We will see how increasing the MODOPS will impact ARK's OC performance. First, we will consider $\mathbf{evks}$ to be pre-loaded on-chip. 

Based on Figure~\ref{fig:arkfreqall}, at low bandwidths, the HKS runtime for different MODOPS is nearly identical, as they are bandwidth limited and do not benefit from increased computational throughput. 
However, at higher bandwidths, HKS becomes compute bound, resulting in an increasing gap between the HKS runtime at different MODOPS. As mentioned earlier, in ARK's OC implementation, off-chip data movement is entirely masked by computation at 128GB/s. We will call this point ``\textit{ARK's saturation point}". 
At 128GB/s, the design is no longer limited by bandwidth. Doubling the MODOPS reduces the HKS runtime. Therefore, as shown in Figure~\ref{fig:arkfreqall}, saturation performance can be achieved with less bandwidth (12.8GB/s) with 2$\times$ MODOPS, saving 10$\times$ on-chip bandwidth.

Table~\ref{tab:ark_modops} outlines the required bandwidth and computational throughput for DC and MP to match ARK's saturation point. Maintaining the same MODOPS as OC necessitates at least 4.26$\times$ and 10$\times$ higher bandwidth for DC and MP, respectively.

\begin{figure}[t!]
  \centering
  \includegraphics[width=1\linewidth, scale=1]{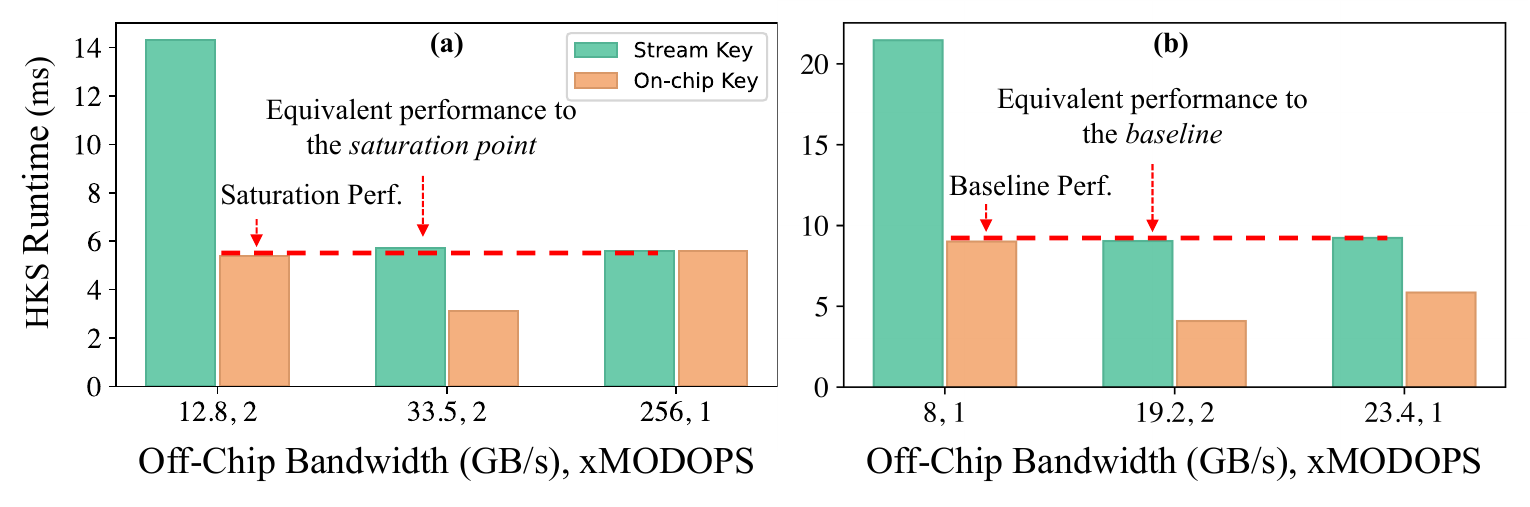}
  \vspace{-1.75em}
  \caption{ARK's configuration for equivalent performance to the baseline and saturation point with OC and streaming evks.}
  \label{fig:arkkeyfreq}
    \vspace{-1.5em}
\end{figure}

Figure \ref{fig:arkkeyfreq} shows the required MODOPS and bandwidth to get the same performance as ``\textit{ARK's saturation point}" and the ``\textit{baseline}" with OC while streaming the $\mathbf{evks}$ on-chip and having a 32MB on-chip memory. We previously showed that 12.8GB/s bandwidth with 2$\times$MODOPS has the same performance as ARK's saturation point, with $\mathbf{evks}$ being on-chip. To get the same performance while streaming the $\mathbf{evks}$, 2.6$\times$ more bandwidth with the same MODOPS (2$\times$) is required, saving 12.25$\times$ on-chip memory. However, with 1$\times$MODOPS, 20$\times$ more bandwidth is required to get the saturation performance. At 256GB/s the design is completely compute bound; therefore, we can find a better balance between bandwidth and MODOPS by increasing the MODOPS and decreasing the bandwidth. This intuition also applies to the \textit{baseline}. As Figure~\ref{fig:arkkeyfreq}(b) shows, by doubling the MODOPS we can achieve the same performance as 1$\times$MODOPS and save 1.2$\times$ bandwidth.  

\section{Related Work}
\noindent\textbf{High-Performance CPU Implementations:}
Many software libraries exist that support HE~\cite{lattigo, OpenFHE, sealcrypto, cheon2017homomorphic}. Notably, Lattigo \cite{lattigo}, OpenFHE \cite{OpenFHE}, and HEAAN \cite{cheon2017homomorphic} support CKKS bootstrapping and FHE. However, due to limited computational power, some applications like deep neural network inference, remain impractical on pure CPU implementations. For instance, state-of-the-art ResNet models \cite{pmlr-v162-lee22e, kim2023hyphen, 10089847} can take upwards of one hour $\textit{per}$ inference, limiting their applicability to infrequent, low arrival rate applications \cite{Garimella_2023, 2023arXiv230704077M}.

\noindent\textbf{Compilers and Dataflow:} Recently, many works have focused on FHE compiler development~\cite{dathathri2018chet, dathathri2020eva, viand2023heco,  ebel2023orion, mad, cowan2021porcupine, ConcreteML, lee2022hecate, malik2023coyote}, particularly for private neural inference. Tools such as CHET \cite{dathathri2018chet} and EVA \cite{dathathri2020eva} automatically map and optimize common neural network layers for FHE, managing parameter selection and ciphertexts implicitly. HECO~\cite{viand2023heco} is an end-to-end compiler that converts high-level programs into secure FHE circuits, allowing non-experts to develop secure and efficient FHE applications. Porcupine~\cite{cowan2021porcupine} and Coyote~\cite{malik2023coyote} generate vectorized HE code for small HE kernels, with Coyote~\cite{malik2023coyote} optimizing data layout to minimize the number of rotations. $\mathsf{Orion}$ \cite{ebel2023orion} utilizes double-hoisting \cite{doublehoist} to enhance CPU latencies. MAD \cite{mad} highlights how caching techniques and dataflow optimizations can yield high performance bootstrapping implementations despite small on-chip memories (1 to 32 MB). While MAD~\cite{mad} presents a dataflow similar to our DC approach, our OC strategy, using 32 and 392MB of on-chip memory, further increases the arithmetic intensity of key-switching by leveraging on-chip data reuse.


\noindent\textbf{GPU/FPGA Acceleration:} 
GPUs offer performance gains with their numerous parallel compute units and high off-chip bandwidth. Previous studies explore GPU acceleration of HE operations~\cite{al2018high, fan2022tensorfhe, Kim_2020, Jung_Kim_Ahn_Cheon_Lee_2021, kaustubh2023gme, shivdikar2022accelerating}. Jung~\cite{Jung_Kim_Ahn_Cheon_Lee_2021} was the first the support CKKS on GPU, suggesting optimizations like \textit{kernel fusing} to reduce on-chip memory requirements. Still, the lack of native modular arithmetic support and limited on-chip memory yields subpar performance compared to modern ASIC solutions \cite{samardzic2022craterlake, kim2022bts, kim2022ark, kimsharp2023}.

Another approach is to use FPGAs~\cite{riazi2020heax, 10070984, 10070953, gener2021fpga, zhu2023fxhenn}. While FPGAs may not provide the same level of performance as ASIC solutions, their re-programmability is advantageous as FHE algorithms evolve. HEAX~\cite{riazi2020heax} is an FPGA-based accelerator for FHE CKKS. Recently, FAB \cite{10070953} and Poseidon~\cite{10070984} proposed FPGA-based accelerators to support bootstrapping. FAB~\cite{10070953} proposes several dataflow optimizations, alongside an FPGA implementation, geared towards increasing on-chip data reuse and minimizing unnecessary DRAM accesses. 
\\
\textbf{ASIC Acceleration:}
Cheetah~\cite{reagen2021cheetah} was the first to present a large-scale ASIC with custom logic and show that the overheads of HE could be overcome with hardware acceleration.
$\mathtt{F1}$~\cite{samardzic2021f1} presented a programmable ASIC accelerator for FHE but was tailored to small parameter sets with limited bootstrapping support, making it inefficient for $\textit{deep}$ neural network computations. Since then, many works \cite{samardzic2022craterlake, kim2022bts, kim2022ark, kimsharp2023} proposed ASIC solutions targeting bootstrapping and FHE. CraterLake \cite{samardzic2022craterlake} and BTS \cite{kim2022bts} were the first ASIC accelerators supporting bootstrapping, but featured underutilized compute units despite allocating 1TB of off-chip bandwidth. Later, ARK \cite{kim2022ark} proposed a minimal key-switching strategy to address memory bottlenecks, yet still required a large 512MB on-chip scratchpad. Recently, SHARP \cite{kimsharp2023} characterized the effect of smaller moduli size on neural network accuracy, demonstrating the feasibility of using smaller 36-bit RNS moduli to reduce compute unit size, on-chip memory requirements, and off-chip bandwidth pressure.

\section{Conclusion}
In this paper, we propose three distinct approaches for implementing the hybrid key-switching (HKS) algorithm that differ in the sequence of instructions and data reuse strategies. The large amount of $\mathbf{evks}$ and generated intermediate data (up to 1.5GB) puts pressure on implementing FHE applications. We introduce a novel dataflow, named Output-Centric (OC), that reduces the off-chip data movement and increases data reuse. Through our evaluations using the RPU accelerator with various benchmarks, our OC dataflow achieves up to 4.16$\times$ speedup over a naive Max-Parallel (MP) implementation of HKS. Additionally, we can save 12.25$\times$ of on-chip SRAM by storing the $\mathbf{evks}$ off-chip and save 3.3$\times$ bandwidth with up to 2.4$\times$ more arithmetic intensity  compared to a MP on-chip implementation.

\section*{Acknowledgment}
Support for this work was provided in part by NSF CAREER award \#2340137.
This work was supported in part by Graduate Assistance in Areas of National Need (GAANN) and
funding from DARPA, under the Data Protection in Virtual Environments (DPRIVE) program, contract HR0011-21-9-0003. 
Reagen and Ebel received generous support from the NY State Center for Advanced Technology in Telecommunications (CATT) and a gift award from Google. 
We also express our gratitude to Duality, especially David Bruce Cousins, Ahmad Al Badawi, and Yuriy Polyakov as well as Andrew G. Schmidt from AMD for their valuable insights and support. 
The views, opinions, and/or findings expressed are those of the authors and do not necessarily reflect the views of sponsors.

\bibliographystyle{ieeetr}
\bibliography{refs}

\end{document}